\documentclass[letterpaper,10pt,prd%,preprint,%,twocolumn
]{revtex4-1}
\usepackage[utf8]{inputenc}\usepackage{newunicodechar}
\newunicodechar{∂}{\partial}
\newunicodechar{∀}{\forall}
\newunicodechar{×}{\times}
\newunicodechar{α}{\alpha}
\newunicodechar{β}{\beta}
\newunicodechar{γ}{\gamma}
\newunicodechar{δ}{\delta}
\newunicodechar{ε}{\epsilon}
\newunicodechar{η}{\eta}
\newunicodechar{φ}{\varphi}
\newunicodechar{ϕ}{\phi}
\newunicodechar{λ}{\lambda}
\newunicodechar{ψ}{\psi}
\newunicodechar{ρ}{\rho}
\newunicodechar{ω}{\omega}
\newunicodechar{τ}{\tau}
\newunicodechar{θ}{\theta}
\newunicodechar{μ}{\mu}
\newunicodechar{ν}{\nu}
\newunicodechar{π}{\pi}
\newunicodechar{σ}{\sigma}
\newunicodechar{ξ}{\xi}
\newunicodechar{Γ}{\Gamma}
\newunicodechar{Ψ}{\Psi}
\newunicodechar{Ω}{\Omega}
\newunicodechar{Σ}{\Sigma}
\newunicodechar{Θ}{\Theta}
\newunicodechar{Λ}{\Lambda}

\usepackage{graphicx}
\usepackage{amsmath}
\usepackage{citesort}
\usepackage{amsfonts}
\usepackage{amssymb}
\usepackage{amsmath,mathtools}
\usepackage{latexsym}
\usepackage{ntheorem}
\usepackage{url}
\usepackage{microtype}
\usepackage{mathtools}
\usepackage{stmaryrd}
\usepackage{bbm}
\usepackage{color,soul}
\usepackage{tensor}
\usepackage{enumerate}
\usepackage{setspace}
\usepackage{verbatim}
\usepackage{pdfpages}

\newcommand{\p}[1]{P\indices{#1}}

\newcounter{mnotecount}[section]

\renewcommand{\themnotecount}{\thesection.\arabic{mnotecount}}
\newcommand{\mnote}[1]%{}
{\protect{\stepcounter{mnotecount}}$^{\mbox{\footnotesize
$%\!\!\!\!\!\!\,
\bullet$\themnotecount}}$ \marginpar{%\color{red}%
\raggedright\tiny\em
$\!\!\!\!\!\!\,\bullet$\themnotecount: #1} }

\newcommand{\jlcax}[1]{}
\newcommand{\eean}{\nonumber\end{eqnarray}}

\newcommand{\kk}[1]{}%{\mnote{{\bf If we consider the KK case:} #1}}

\newcommand{\beq}{\begin{equation}}

\newcommand{\FS}       %{F_1} %
                  {F}
\newcommand{\HS} %{F_2}
       {H_{\mbox{\scriptsize volume}}}

{\ptc{this should be removed in the oberwolfach version}}%

\newcommand{\eeal}[1]{\label{#1}\end{eqnarray}}
\newcommand{\bed}{\begin{deqarr}}
\newcommand{\eed}{\end{deqarr}}
\newcommand{\bedl}[1]{\begin{deqarr}\label{#1}}
\newcommand{\eedl}[2]{\arrlabel{#1}\label{#2}\end{deqarr}}

\newcommand{\bel}[1]{\begin{equation}\label{#1}}
\newcommand{\bea}{\begin{eqnarray}}
\newcommand{\bean}{\begin{eqnarray}\nonumber}
\newcommand{\beal}[1]{\begin{eqnarray}\label{#1}}
\newcommand{\eea}{\end{eqnarray}}

\newcommand{\nn}{\nonumber}

\newcommand{\be}{\begin{equation}}
\newcommand{\eeq}{\end{equation}}
\newcommand{\ee}{\end{equation}}
\newcommand{\beqa}{\begin{eqnarray}}
\newcommand{\eeqa}{\end{eqnarray}}
\newcommand{\beqan}{\begin{eqnarray*}}
\newcommand{\eeqan}{\end{eqnarray*}}
\newcommand{\ba}{\begin{array}}
\newcommand{\ea}{\end{array}}

\newcommand{\mcM}{{\mycal M}}

\newcommand{\dx}{\,dx}

\newcommand{\warn}[1]%{}%{}
{\protect{\stepcounter{mnotecount}}$^{\mbox{\footnotesize
$%\!\!\!\!\!\!\,
\bullet$\themnotecount}}$ \marginpar{%\color{red}%
\raggedright\tiny\em
$\!\!\!\!\!\!\,\bullet$\themnotecount: {\bf Warning:} #1} }

\newcommand{\N}{\mathbb N}

\newcommand{\eq}[1]{(\ref{#1})}

\newcommand{\ptc}[1]{\mnote{{\bf ptc:}#1}}

\newcommand{\beqar}{\begin{deqarr}}
\newcommand{\eeqar}{\end{deqarr}}

\newcommand{\beaa}{\begin{eqnarray*}}
\newcommand{\eeaa}{\end{eqnarray*}}

\DeclareFontFamily{OT1}{rsfs}{}
\DeclareFontShape{OT1}{rsfs}{m}{n}{ <-7> rsfs5 <7-10> rsfs7 <10-> rsfs10}{}
\DeclareMathAlphabet{\mycal}{OT1}{rsfs}{m}{n}

\usepackage[table]{xcolor}
\definecolor{darkblue}{rgb}{0,0,.5}

{\catcode `\@=11 \global\let\AddToReset=\@addtoreset}
\AddToReset{equation}{section}

{\catcode `\@=11 \global\let\AddToReset=\@addtoreset}
\AddToReset{figure}{section}

\newcommand{\D}{\mathrm{d}}
\renewcommand{\dx}{\mathrm{d}x}

\renewcommand{\N}[1]{\mathcal{N}\indices{#1}}
\newcommand{\Ni}[1]{(\mathcal{N}^{-1})\indices{#1}}
\renewcommand{\p}[1]{P\indices{#1}}

\begin{document}

\title{Vacuum space-times with controlled singularities and without symmetries}%
\preprint{UWThPh-2015-16}
\author{Piotr T. Chru\'sciel}
\email[Email ]{piotr.chrusciel@univie.ac.at}
\homepage[URL ]{homepage.univie.ac.at/piotr.chrusciel}
\affiliation{Faculty of Physics and Erwin Schr\"odinger Institute\\ Boltzmanngasse 5, 1090 Vienna, Austria}

\author{Paul Klinger}
\email[Email ]{paul.klinger@univie.ac.at}
\affiliation{Faculty of Physics and Erwin Schr\"odinger Institute\\ Boltzmanngasse 5, 1090 Vienna, Austria}

\begin{abstract}
We show existence of a family of four-dimensional vacuum space-times with asymptotically velocity-dominated singularities and without symmetries. The solutions are obtained using {Fuchsian methods} and are parameterised by several free functions of all space coordinates which control their asymptotic expansion.
\end{abstract}
\maketitle
%\hspace{2.1em}

\noindent
A key question in mathematical general relativity is the understanding of the dynamics of the gravitational field when singularities are approached. It has been conjectured~\cite{BKL} that the behaviour of the metric will be rather involved, exhibiting a complicated ``mixmaster'' behaviour reminiscent of that encountered in oscillating Bianchi models~\cite{Ringstroem2}. However, in spite of many interesting studies (cf., e.g., \cite{HeinzleUggla,DamourBuyl,Berger:2000uf,BGIMW} and references therein), the issue remains wide open. In fact, except for the finite-dimensional families of \cite{Ringstroem2,Berger:2000uf}, all remaining four-dimensional \emph{vacuum} singularities rigorously constructed so far~\cite{RingstroemSCC,Isenberg:2002jg} exhibit ``asymptotically velocity-dominated behaviour''. Moreover, all four dimensional vacuum examples with well understood dynamical behaviour near a singularity  involve  metrics with at least a one-dimensional isometry group. The purpose of this note is to point out that one can use the approach developed in~\cite{DamourBuyl} to construct a family of vacuum examples
 with velocity-dominated asymptotics and
 without any symmetries. (See~\cite{AnderssonRendall,Anguige:1999gg,Anguige:1999gh} for non-vacuum four-dimensional examples, and~\cite{DHRW} for vacuum higher-dimensional ones. Further references can be found in~\cite{JimMillenium}.)

It is clear from the ansatz below that the solutions we construct are highly non-generic. While they do not tell us anything about what happens in the generic case, they provide the largest class known so far of vacuum four-dimensional space-times with controlled behaviour as the singularity is approached.

As such, we consider metrics of the form
\begin{equation}\label{10II15.1}
g=-e^{-2\sum_{a=1}^3  β^a}\D τ^2+\sum_{a=1}^3  e^{-2β^a}\N{^a_i}\N{^a_j}\dx^i\dx^j\,,
\end{equation}
with $β^a$ and $\N{^a_i}$, $i,a\in\{1,2,3\}$,  depending on all coordinates $τ$, $x^i$ and behaving asymptotically as
\begin{equation}\label{10II15.2}
β^a=β_\circ^a+τp_\circ^a+O(e^{-τν})\quad\text{and}\quad
\N{^a_i}=:δ^a_i+ \N{_s^a_i}=δ^a_i+O(e^{-τν})
 \,,
\end{equation}
where $ν$ is a positive constant and $\N{_s^a_i}=0$ for $a\ge i$, while the $β_\circ^a$'s and  $p_\circ^a$'s depend only upon space coordinates. In fact we have the more precise expansions
\bea\label{16VI15.4}
    \N{_s^1_2}
     &= &
     -\frac{\p{_\circ^2_1}e^{-2(β_\circ^2-\beta_\circ^1)}}{2(p_\circ^2-p_\circ^1)}
      e^{-τ(2p_\circ^2-2p_\circ^1 )}+O(e^{-τ(2p_\circ^2-2p_\circ^1 +ν)})
     \,,
\\
 \label{16VI15.5}
    \N{_s^2_3}
     & = &
       -\frac{\p{_\circ^3_2}e^{-2(β_\circ^3-β_\circ^2)}}{2(p_\circ^3-p_\circ^2)}
         e^{-τ(2p_\circ^3-2p_\circ^2 )}
      +O(e^{-τ(2p_\circ^3-2p_\circ^2 +ν)})
      \,,
\\
    \N{_s^1_3}
     &= &
      e^{-2(β_\circ^3-β_\circ^1)}\bigg(\p{_\circ^3_1}
        -\frac{\p{_\circ^2_1}\p{_\circ^3_2}}{2p_\circ^3-2p_\circ^2}\bigg)\frac{1}{2p_\circ^3
        -2p_\circ^1}
       e^{-τ(2p_\circ^3-2p_\circ^1 )}
     \nn
\\
     &&
      +O(e^{-τ(2p_\circ^3-2p_\circ^1 +ν)})
     \,,
\label{16VI15.6}
\eea
where the functions $\{P\indices{_\circ^i_a}\}_{1\le a < i\le 3}$ depend only on space coordinates.

Our solutions are parametrised by \emph{freely} prescribable analytic functions $β_\circ^2$, $β_\circ^3$ and $P\indices{_\circ^2_1}$
of \emph{all} space coordinates,
as well as two analytic functions, $p_\circ^2$ and $p_\circ^3$  depending on \emph{all} space coordinates, which are \emph{free except for the inequalities}
\bel{13VI15.1}
0<p_\circ^2<(\sqrt{2}-1)p_\circ^3\,.
\ee
The remaining functions $p_\circ^1$, $β^1_{\circ,3}$,  $\p{_\circ^3_2 }$ and $\p{_\circ^3_1 }$ are then determined by the asymptotic constraint equations:
\begin{align}
 \label{final_my_p_cond}
    %  p_\circ^3>0\,,\qquad %0<p_\circ^2<(\sqrt{2}-1)p_\circ^3\quad\text{and}\quad
    p_\circ^1 &=-\frac{p_\circ^2 p_\circ^3}{p_\circ^2+p_\circ^3}\,,
\\
β^1_{\circ,3}
 &=-(p_\circ^2+p_\circ^3)^{-1}(p_{\circ,3}^2+p_{\circ,3}^1+β_{\circ,3}^2(p_\circ^1
 +p_\circ^3)+β^3_{\circ,3}(p_\circ^1+p_\circ^2))\,,
 \\
 \p{_\circ^3_2_{,3}}&=2\left(G_{2c}p_{\circ,2}^c+β_{\circ,2}^d p_\circ^f G_{df}\right)\,,
\\
 \label{28X14.1}
\p{_\circ^3_1_{,3}}&=-\p{_\circ^2_1_{,2}}+2\left(G_{1c}p_{\circ,1}^c+β^d_{\circ,1}p_\circ^f G_{df}\right)
 \,.
\end{align}
Here the $3\times 3$ matrix $G^{ab}=(2 δ^{ab} -1)/2$ and its inverse  $G_{ab}=-\sum_{c\neq d}δ^c_aδ^d_b$ can be explicitly written as
\[
(G_{ab})=\begin{pmatrix*}[r]0&-1&-1\\-1&0&-1\\-1&-1&0\end{pmatrix*}\quad\text{and}\quad
(G^{ab})=\frac{1}{2}\begin{pmatrix*}[r]1&-1&-1\\-1&1&-1\\-1&-1&1\end{pmatrix*}\,.
\]

These solutions arise as follows: The vacuum Einstein equations for a metric of the form \eq{10II15.1} can be encoded in the Hamiltonian~\cite{DamourBuyl}
\begin{widetext}
\begin{equation}\label{ham_iwa_full}\begin{split}
H=&\frac{1}{4}G^{ab}π_a π_b+\sum_{a<b}\frac{1}{2}(\p{^j_a}\N{^b_j})^2 e^{-2(β^b-β^a)}+\hspace{-0.9em}\sum_{a\neq b\neq c\neq a}\hspace{-0.9em}\frac{1}{4}(C\indices{^a_b_c})^2
 e^{-4β^a }\\
&-\sum_{a}\bigg[
-2(β^a_{,a})^2-2β^a_{,a,a}
+\sum_{b}\bigg(-2(C\indices{^b_a_b})^2
-4C\indices{^b_b_a}β^a_{,a}+4β^b_{,a}β^a_{,a}-(β^b_{,a})^2
\\&\hspace{3.5em}-2C\indices{^b_a_b}β^b_{,a}
+2β^b_{,a,a}+2C\indices{^b_a_b_{,a}}
\\&\hspace{3.5em}+\sum_c\bigg(
C\indices{^b_b_a}C\indices{^c_a_c}-β^b_{,a}β^c_{,a}-C\indices{^b_a_c}C\indices{^c_a_b}/2-2C\indices{^b_a_b}β^c_{,a}\bigg)\bigg)
 \bigg]e^{-2\sum_{c\neq a}β^c}
 %+2Λe^{-2\sum_aβ^a}
 \,,
\end{split}\end{equation}
\end{widetext}
with $C\indices{^a_b_c}=\sum_{i,k}2\N{^a_k}\Ni{^i_{[b}}\Ni{^k_{c],i}}$,
 the derivative operator ``$_{,a}$'' is defined as $_{,a}=\Ni{^i_a}∂_i$, and where the $\pi_a$'s are  canonically conjugate to the $\beta^a$'s, while the $ \p{^j_a}$'s are canonically conjugate to the $\N{^a_j}$'s.

It is relatively straightforward, though somewhat tedious, to check that Hamilton's evolution equations with the ansatz \eq{10II15.2}-\eq{final_my_p_cond} verify the hypotheses of the ``Fuchs theorem'' of Choquet-Bruhat~\cite[Appendix~V, p.~636]{YCB:GRbook}.
This gives existence of solutions of the evolution equations.

To show that these also satisfy the constraint equations, a system of evolution equations for the constraints is derived which is homogeneous and also verifies the hypotheses of the Fuchs theorem. As the full constraints approach the asymptotic ones, which vanish,  and the Fuchs theorem guarantees that the only asymptotically vanishing solution to a homogeneous Fuchsian system is identically zero,  the constraints are satisfied.
This establishes existence of vacuum space-times as above.

The question then arises, what are the isometries of the metrics just constructed.
In \cite{DamourHenneauxNicolai} it is asserted that transformations mixing time and space coordinates are prohibited by the choice of lapse and shift and the assumption that the singularity is approached as $τ\to\infty$.
%Presumably this should follow from the resulting equations
%%
%\begin{align}
%g_{ij}\frac{∂x^i}{∂\tilde{τ}}\frac{∂x^j}{∂y^k}&=\det g \frac{∂τ}{∂\tilde{τ}}\frac{∂τ}{∂y^k}
% \;,
% \label{coord_shift}
%\\
% -\det g \left(\frac{∂τ}{∂\tilde{τ}}\right)^2+g_{ij}\frac{∂x^i}{∂\tilde{τ}}\frac{∂x^j}{∂\tilde{τ}}&=-\det\left(-\det g\left(\frac{∂τ}{∂y^k}\right)^2δ_{kl}+g_{ij}\frac{∂x^i}{∂y^k}\frac{∂x^j}{∂y^l}\right)
% \,,
% \label{coord_lapse}
%\end{align}
%%
%(assuming a transformation $τ,x^i\to \tilde{τ}(τ,x^j),y^i(τ, x^j)$).
While this is plausible,  the assertion is not clear and we have not been able to provide a proof. Assuming nevertheless that all isometries do indeed preserve the $\tau$-slicing, Killing vectors $X$ of $g$ should have vanishing $\tau$ component: $X(\tau)=0$. Under this last condition, a calculation shows that generic choices of the free functions above only lead to trivial Killing vector fields. More generally, one can show that generic metrics in our class do not have any isometries that preserve the $\tau$-slicing of the space-time $(\mcM,g)$.

We have
\[
 R_{αβγδ}R^{αβγδ}=\left(\frac{16e^{4(β_\circ^1+β_\circ^2+β_\circ^3)}\big(p_\circ^2 p_\circ^3\big)^2  }{(p_\circ^2+p_\circ^3)^2}
       \big((p_\circ^2)^2+p_\circ^2p_\circ^3+(p_\circ^3)^2\big)+O(e^{-\nu\tau})
 \right)e^{\tau 4(p_\circ^1+p_\circ^2+p_\circ^3)}
 \,,
\]
which shows that the curvature tensor grows uniformly without bounds on all causal curves in the space-times constructed above, since the product $p_\circ^2 p_\circ^3$ has no zeros.

We note that our preliminary attempts to find  an ansatz for higher dimensional solutions that is compatible with the Fuchs theorem and has no symmetries have not been successful.

It should be pointed out that our considerations are unaffected by the presence of a cosmological constant. Indeed, all equations above remain unchanged, except for the addition of a term $ 2\Lambda e^{-2\sum_a \beta^a}$ in (11), with $\Lambda$ influencing only lower order terms in the asymptotic expansions of the solutions.

Details of the analysis outlined above can be found in~\cite{Klinger1}.

\begin{acknowledgments}
PTC acknowledges many useful discussions with Jim Isenberg. Supported in part by Narodowe Centrum Nauki under the grant DEC-2011/03/B/ST1/02625.
\end{acknowledgments}

\bibliography{../references/hip_bib,%
../references/reffile,%
../references/newbiblio,%
../references/newbiblio2,%
../references/bibl,%
../references/howard,%
../references/bartnik,%
../references/myGR,%
../references/newbib,%
../references/Energy,%
../references/netbiblio}
\end{document}